\documentclass[aps,prl,twocolumn,superscriptaddress,showpacs,preprintnumbers,amsmath,amssymb]{revtex4}
\usepackage{epsfig}
\usepackage{multirow}

\usepackage{graphicx} 
\usepackage{dcolumn} 
\usepackage{rotating}
\usepackage{adjustbox}
\usepackage{threeparttable}
\usepackage[colorlinks,linkcolor=red,anchorcolor=green,citecolor=blue]{hyperref}

\newcommand{\gisr}{\gamma_{\rm ISR}}
\newcommand{\gh}{\gamma_{\rm h}}
\newcommand{\gl}{\gamma_{\rm l}}
\newcommand{\x}{X(3872)}

\newcommand{\yfos}{\Upsilon(4S)}
\newcommand{\yfis}{\Upsilon(5S)}

\newcommand{\eff}{\varepsilon}
\newcommand{\BR}{{\cal B}}

\newcommand{\pim}{\pi^-}

\newcommand{\kap}{K^+}

\newcommand{\ks}{K_S^0}

\newcommand{\psp}{\psi(2S)}

\newcommand{\jpsi}{J/\psi}

\newcommand{\chicz}{\chi_{c0}}
\newcommand{\chico}{\chi_{c1}}
\newcommand{\chict}{\chi_{c2}}
\newcommand{\chicj}{\chi_{cJ}}
\newcommand{\etac}{\eta_c}
\newcommand{\EE}{e^+e^-}
\newcommand{\MM}{\mu^+\mu^-}
\newcommand{\LL}{\ell^+\ell^-}
\newcommand{\pp}{\pi^+\pi^-}
\newcommand{\kk}{K^+K^-}

\newcommand{\fb}{\rm fb}

\newcommand{\infb}{\rm fb^{-1}}

\newcommand{\gev}{\rm GeV}

\newcommand{\gevcs}{{\rm GeV}/c^2}
\newcommand{\mev}{\rm MeV}
\newcommand{\mevcs}{{\rm MeV}/c^2}
\newcommand{\beq}{\begin{equation}}
\newcommand{\eeq}{\end{equation}}
\newcommand{\bitm}{\begin{itemize}}
\newcommand{\eitm}{\end{itemize}}

\begin{document}

\title{\quad\\[1.0cm] Observation of $e^+e^- \to \gamma\chico$ and search for $e^+e^- \to \gamma \chicz, \gamma \chict,$ and $\gamma\etac$ at $\sqrt{s}$ near 10.6 GeV at Belle}

\noaffiliation
\affiliation{University of the Basque Country UPV/EHU, 48080 Bilbao}
\affiliation{Beihang University, Beijing 100191}
\affiliation{Brookhaven National Laboratory, Upton, New York 11973}
\affiliation{Budker Institute of Nuclear Physics SB RAS, Novosibirsk 630090}
\affiliation{Faculty of Mathematics and Physics, Charles University, 121 16 Prague}
\affiliation{University of Cincinnati, Cincinnati, Ohio 45221}
\affiliation{Deutsches Elektronen--Synchrotron, 22607 Hamburg}
\affiliation{University of Florida, Gainesville, Florida 32611}
\affiliation{Key Laboratory of Nuclear Physics and Ion-beam Application (MOE) and Institute of Modern Physics, Fudan University, Shanghai 200443}
\affiliation{Justus-Liebig-Universit\"at Gie\ss{}en, 35392 Gie\ss{}en}
\affiliation{II. Physikalisches Institut, Georg-August-Universit\"at G\"ottingen, 37073 G\"ottingen}
\affiliation{SOKENDAI (The Graduate University for Advanced Studies), Hayama 240-0193}
\affiliation{Gyeongsang National University, Chinju 660-701}
\affiliation{Hanyang University, Seoul 133-791}
\affiliation{University of Hawaii, Honolulu, Hawaii 96822}
\affiliation{High Energy Accelerator Research Organization (KEK), Tsukuba 305-0801}
\affiliation{J-PARC Branch, KEK Theory Center, High Energy Accelerator Research Organization (KEK), Tsukuba 305-0801}
\affiliation{Forschungszentrum J\"{u}lich, 52425 J\"{u}lich}
\affiliation{IKERBASQUE, Basque Foundation for Science, 48013 Bilbao}
\affiliation{Indian Institute of Technology Bhubaneswar, Satya Nagar 751007}
\affiliation{Indian Institute of Technology Guwahati, Assam 781039}
\affiliation{Indian Institute of Technology Hyderabad, Telangana 502285}
\affiliation{Indian Institute of Technology Madras, Chennai 600036}
\affiliation{Indiana University, Bloomington, Indiana 47408}
\affiliation{Institute of High Energy Physics, Chinese Academy of Sciences, Beijing 100049}
\affiliation{Institute of High Energy Physics, Vienna 1050}
\affiliation{Institute for High Energy Physics, Protvino 142281}
\affiliation{INFN - Sezione di Napoli, 80126 Napoli}
\affiliation{INFN - Sezione di Torino, 10125 Torino}
\affiliation{Advanced Science Research Center, Japan Atomic Energy Agency, Naka 319-1195}
\affiliation{J. Stefan Institute, 1000 Ljubljana}
\affiliation{Institut f\"ur Experimentelle Teilchenphysik, Karlsruher Institut f\"ur Technologie, 76131 Karlsruhe}
\affiliation{Kennesaw State University, Kennesaw, Georgia 30144}
\affiliation{Department of Physics, Faculty of Science, King Abdulaziz University, Jeddah 21589}
\affiliation{Kitasato University, Sagamihara 252-0373}
\affiliation{Korea Institute of Science and Technology Information, Daejeon 305-806}
\affiliation{Korea University, Seoul 136-713}
\affiliation{Kyungpook National University, Daegu 702-701}
\affiliation{LAL, Univ. Paris-Sud, CNRS/IN2P3, Universit\'{e} Paris-Saclay, Orsay}
\affiliation{\'Ecole Polytechnique F\'ed\'erale de Lausanne (EPFL), Lausanne 1015}
\affiliation{P.N. Lebedev Physical Institute of the Russian Academy of Sciences, Moscow 119991}
\affiliation{Faculty of Mathematics and Physics, University of Ljubljana, 1000 Ljubljana}
\affiliation{Ludwig Maximilians University, 80539 Munich}
\affiliation{Luther College, Decorah, Iowa 52101}
\affiliation{Malaviya National Institute of Technology Jaipur, Jaipur 302017}
\affiliation{University of Malaya, 50603 Kuala Lumpur}
\affiliation{University of Maribor, 2000 Maribor}
\affiliation{Max-Planck-Institut f\"ur Physik, 80805 M\"unchen}
\affiliation{School of Physics, University of Melbourne, Victoria 3010}
\affiliation{University of Mississippi, University, Mississippi 38677}
\affiliation{University of Miyazaki, Miyazaki 889-2192}
\affiliation{Moscow Physical Engineering Institute, Moscow 115409}
\affiliation{Moscow Institute of Physics and Technology, Moscow Region 141700}
\affiliation{Graduate School of Science, Nagoya University, Nagoya 464-8602}
\affiliation{Kobayashi-Maskawa Institute, Nagoya University, Nagoya 464-8602}
\affiliation{Universit\`{a} di Napoli Federico II, 80055 Napoli}
\affiliation{Nara Women's University, Nara 630-8506}
\affiliation{National Central University, Chung-li 32054}
\affiliation{National United University, Miao Li 36003}
\affiliation{Department of Physics, National Taiwan University, Taipei 10617}
\affiliation{H. Niewodniczanski Institute of Nuclear Physics, Krakow 31-342}
\affiliation{Nippon Dental University, Niigata 951-8580}
\affiliation{Niigata University, Niigata 950-2181}
\affiliation{Novosibirsk State University, Novosibirsk 630090}
\affiliation{Osaka City University, Osaka 558-8585}
\affiliation{Pacific Northwest National Laboratory, Richland, Washington 99352}
\affiliation{Panjab University, Chandigarh 160014}
\affiliation{Peking University, Beijing 100871}
\affiliation{University of Pittsburgh, Pittsburgh, Pennsylvania 15260}
\affiliation{Punjab Agricultural University, Ludhiana 141004}
\affiliation{Theoretical Research Division, Nishina Center, RIKEN, Saitama 351-0198}
\affiliation{University of Science and Technology of China, Hefei 230026}
\affiliation{Seoul National University, Seoul 151-742}
\affiliation{Showa Pharmaceutical University, Tokyo 194-8543}
\affiliation{Soongsil University, Seoul 156-743}
\affiliation{University of South Carolina, Columbia, South Carolina 29208}
\affiliation{Stefan Meyer Institute for Subatomic Physics, Vienna 1090}
\affiliation{Sungkyunkwan University, Suwon 440-746}
\affiliation{School of Physics, University of Sydney, New South Wales 2006}
\affiliation{Department of Physics, Faculty of Science, University of Tabuk, Tabuk 71451}
\affiliation{Tata Institute of Fundamental Research, Mumbai 400005}
\affiliation{Department of Physics, Technische Universit\"at M\"unchen, 85748 Garching}
\affiliation{Toho University, Funabashi 274-8510}
\affiliation{Department of Physics, Tohoku University, Sendai 980-8578}
\affiliation{Earthquake Research Institute, University of Tokyo, Tokyo 113-0032}
\affiliation{Department of Physics, University of Tokyo, Tokyo 113-0033}
\affiliation{Tokyo Institute of Technology, Tokyo 152-8550}
\affiliation{Tokyo Metropolitan University, Tokyo 192-0397}
\affiliation{Virginia Polytechnic Institute and State University, Blacksburg, Virginia 24061}
\affiliation{Wayne State University, Detroit, Michigan 48202}
\affiliation{Yamagata University, Yamagata 990-8560}
\affiliation{Yonsei University, Seoul 120-749}
\author{S.~Jia}\affiliation{Beihang University, Beijing 100191} 
\author{X.~L.~Wang}\affiliation{Key Laboratory of Nuclear Physics and Ion-beam Application (MOE) and Institute of Modern Physics, Fudan University, Shanghai 200443} 
\author{C.~P.~Shen}\affiliation{Beihang University, Beijing 100191} 
 \author{C.~Z.~Yuan}\affiliation{Institute of High Energy Physics, Chinese Academy of Sciences, Beijing 100049} 
  \author{I.~Adachi}\affiliation{High Energy Accelerator Research Organization (KEK), Tsukuba 305-0801}\affiliation{SOKENDAI (The Graduate University for Advanced Studies), Hayama 240-0193} 
  \author{H.~Aihara}\affiliation{Department of Physics, University of Tokyo, Tokyo 113-0033} 
  \author{S.~Al~Said}\affiliation{Department of Physics, Faculty of Science, University of Tabuk, Tabuk 71451}\affiliation{Department of Physics, Faculty of Science, King Abdulaziz University, Jeddah 21589} 
  \author{D.~M.~Asner}\affiliation{Brookhaven National Laboratory, Upton, New York 11973} 
  \author{H.~Atmacan}\affiliation{University of South Carolina, Columbia, South Carolina 29208} 
  \author{V.~Aulchenko}\affiliation{Budker Institute of Nuclear Physics SB RAS, Novosibirsk 630090}\affiliation{Novosibirsk State University, Novosibirsk 630090} 
% ...
\collaboration{The Belle Collaboration}


\begin{abstract}

Using data samples of 89.5~fb$^{-1}$, 711.0~fb$^{-1}$, and 121.4~fb$^{-1}$
collected with the Belle detector at the KEKB asymmetric-energy $e^+e^-$ collider
at center-of-mass energies 10.52 GeV, 10.58 GeV, and 10.867 GeV, respectively,
 we study the exclusive reactions
$\EE\to\gamma\chicj$ ($J = 0,~1,~2$) and $\EE\to\gamma\etac$.
 A significant $\gamma \chi_{c1}$ signal is
observed for the first time at $\sqrt{s}=10.58~\gev$
 with a significance of $5.1\sigma$ including systematic uncertainties.
 No significant excesses for
$\gamma \chi_{c0}$, $\gamma \chi_{c2}$, and $\gamma \etac$ final states are found,
and we set 90\% credibility level upper limits on the Born cross sections ($\sigma_{\rm B}$) at 10.52 GeV, 10.58 GeV, and 10.867~GeV.
Together with cross sections measured by BESIII at lower center-of-mass energies, the energy dependency of
$\sigma_{\rm B}(\EE\to\gamma\chico)$ is obtained.

\end{abstract}


\maketitle


The production of heavy quark pairs in high-energy lepton collisions is
described well by perturbative Quantum Chromodynamics (pQCD). Yet a description of these pairs forming quarkonium---charmonium or bottomonium---is theoretically challenging.
Quarkonium formation is governed by nonperturbative long-distance effects~\cite{74.2981}.
Nonrelativistic Quantum Chromodynamics (NRQCD) factorization was used to compute the cross section for several processes, including the double-charmonium production cross section~\cite{96.092001,98.092003}, $\EE \to \gamma \chicj$ ($J=0,~1,~2$)~\cite{D78.074022,D81.034028, D80.114014,JHEP1410,1712.06165} and $\EE \to \gamma \etac$~\cite{D81.034028, D80.114014,B186.475,1801.091} at $B$ factories with relativistic and higher-order corrections included.

Electromagnetic quarkonium production is relatively simpler than other production mechanisms,
and therefore it serves as a good testing ground for such NRQCD predictions.
The BESIII experiment measured $\EE \to \gamma \chicj$ cross sections at $\sqrt{s} =
4.01~\gev$, $4.23~\gev$, $4.26~\gev$, and $4.36~\gev$ and $\EE \to \gamma \etac$ cross section
at the same energies and additionally at 4.42 GeV and 4.60 GeV~\cite{C39.041001,D96.051101}.
At none of the individual energy points does the statistical significance for production of $\chi_{cJ}$ or $\eta_c$ exceed $3\sigma$, and when the data from all energy points are combined, the statistical significances for $\chi_{c1}$, $\chi_{c2}$, and $\eta_c$ are $3.0\sigma$, $3.4\sigma$, and greater than $3.6\sigma$, respectively.

In addition, the BESIII experiment reported evidence for $\x$ production via $\EE \to \gamma \x$~\cite{1310.4101}.
 Precise and unambiguous measurement of $\EE \to \gamma \chicj$ and $\gamma
\etac$ is useful for understanding $C$-even quarkonia and exotic $XYZ$
particles~\cite{D90.034020,1310.8597,D79.094504}, e.g., $\x$.

In this paper, we report cross-section measurements for the exclusive reactions $\EE \to \gamma \chicj$ and $\gamma \etac$ with data recorded at $\sqrt{s} \sim 10.6~\gev$ by the Belle experiment at the KEKB asymmetric-energy
$\EE$ collider~\cite{KEKB1,KEKB2}.
The data used in this analysis corresponds to $89.5~\infb$ of integrated luminosity at $10.52~\gev$, referred to as the continuum sample; $711~\infb$ at $10.58~\gev$, referred to as the $\yfos$ sample; and $121.4~\infb$ at $10.867~\gev$, referred to as the $\yfis$ sample.


The Belle detector~\cite{Belle1,Belle2} is a large solid-angle magnetic spectrometer that consists of a silicon vertex
 detector, a 50-layer central drift chamber (CDC), an array of aerogel threshold Cherenkov counters
(ACC), a barrel-like arrangement of time-of-flight scintillation counters (TOF), and an
 electromagnetic calorimeter comprised of CsI(Tl) crystals (ECL) located inside a superconducting
 solenoid coil that provides a $1.5~\hbox{T}$ magnetic field. An iron flux-return yoke instrumented
 with resistive plate chambers (KLM) located outside the coil is used to detect $K^{0}_{L}$ mesons
and to identify muons. The Belle detector is described in detail elsewhere~\cite{Belle1,Belle2}.

We determined event-selection criteria using a large sample of Monte Carlo (MC) signal events (100k) for $e^+e^- \to \gamma \chicj$ and $\gamma\etac$ at $\sqrt{s}$ = 10.52, 10.58, and $10.867~\gev$ generated with EvtGen~\cite{EVTGEN}.
In the generator, the polar angle of the transition photon in the $\EE$ C.M. system ($\theta_\gamma$) is distributed according to $(1 + \cos^2\theta_\gamma)$ for $\gamma \chicz$ and $\gamma \etac$ production, and $(1 + 0.63\cos^2\theta_\gamma)$ for $\gamma \chico$ production~\cite{DPNU}.
No definite model exists for the distribution of $\theta_\gamma$ in $\gamma \chict$ production because the combination of tensor-meson production and $\gamma$ emission is theoretically complicated and requires experimental input. So we model the production of this channel as evenly distributed in phase space and account for differences from $(1 \pm {\rm cos}^2\theta_\gamma)$ distributions as systematic uncertainties.

Corrections due to initial-state radiation (ISR) are taken into account in all studied channels, where we assume $\sigma(\EE \to \gamma \chicj/\etac)
\sim 1/s^{n}$ in the calculation of the radiative-correction factor. 
The values of $n$, determined from Refs.~\cite{1712.06165,1801.091}, are 1.4 for $\chicz$, 2.1 for $\chico$, 2.4 for $\chict$, and 1.3 for $\etac$ in the predictions of next-to-leading order (NLO) QCD, and 1.4 for $\etac$ in leading order (LO) QCD.
Possible sources of background events from $\Upsilon(nS) \to B\bar B$~($n$ = 4, 5), $\yfis \to B^{(*)}_s\bar{B}^{(*)}_s$,
 and $\EE \to q\bar{q}~(q = u, d, s, c)$ are checked with a MC sample four times larger than the data sample, and are also generated with EvtGen~\cite{EVTGEN}.
GEANT3~\cite{GEANT3} is used to simulate the detector response to all MC events.
The $\chicj$ candidates are reconstructed from their decays to $\gamma\jpsi$ with $\jpsi \to \MM$, and the $\etac$ candidates are reconstructed from five hadronic decays into $\ks\kap\pim$, $\pp\kk$, $2(\pp)$, $2(\kk)$, and $3(\pp)$~\cite{conjugate}.


We define a well-reconstructed charged track as having impact parameters with respect to the nominal interaction point of less than $0.5~{\rm cm}$ and $4~{\rm cm}$ perpendicular to and along the beam direction, respectively.
For a $\EE\to\gamma\chicj$ candidate event, we require the number of well-reconstructed charged
tracks, $N_{\rm trk}$, to be two, and the net charge be zero. For
 $\EE\to\gamma\etac$, we require $N_{\rm trk} = 6$ for the $3(\pp)$ final state and $N_{\rm trk} = 4$ for
 the other final states, also with a zero net charge.
For the particle identification (PID) of a well-reconstructed charged track, information from different
detector subsystems, including specific ionization in the CDC, time measurement in the TOF, and the
 response of the ACC, is combined to form a likelihood ${\mathcal L}_i$~\cite{A494.402} for particle species $i$.
 Tracks with $R_{K}=\mathcal{L}_{K}/(\mathcal{L}_K+\mathcal{L}_\pi)<0.4$ are
identified as pions with an efficiency of 96\%, while 9\% of kaons are misidentified as pions;
tracks with $R_{K}>0.6$ are identified as kaons with an efficiency of 98\%, while 8\% of pions
are misidentified as kaons.



For muons from $\jpsi\to\MM$, we require at least one of the paired tracks to have $R_{\mu} =
\mathcal{L}_{\mu}/(\mathcal{L}_{\mu}+\mathcal{L}_{K} + \mathcal{L}_{\pi}) > 0.95$; if one track has $R_{\mu}<0.95$, it must have associated hits in the KLM agreeing with the extrapolated trajectory provided by the CDC~\cite{Abashian49169}. The efficiency of muon-pair identification is $94\%$.

Using a multivariate analysis with a neural network~\cite{A559.190} based on two sets of input variables~\cite{Thesis}, a
$\ks$ candidate is reconstructed from a pair of oppositely-charged tracks that are treated as
pions. An ECL cluster with energy higher than $50~\mev$ is treated as a photon candidate if it does
not match the extrapolation of any charged track. The photon with the maximum energy in the $\EE$
 C.M. system is taken as the transition photon. Since there are two photons in the $\EE\to\gamma \chicj$
channel, the transition photon is denoted as $\gh$, and the one with the second highest energy is
denoted as $\gl$ and is taken as the photon from the $\chi_{cJ}\to\gamma J/\psi$ decay.
We require $E(\gl)>300~\mev$ to suppress the backgrounds from fake photons.

If there are more than two photons, to suppress the background from the ISR process $\EE \to \gisr \psp \to \gamma_{{\rm ISR}} \gamma \chicj$, an
 extra photon ($\gamma_{\rm ext}$) besides $\gh$ and $\gl$ is selected, and $M(\gamma_{\rm ext}
\gl \MM) < 3.60 ~\gevcs$ or $M(\gamma_{\rm ext} \gl\MM)>3.78~\gevcs$ is required. This requirement removes 92.2\% and 91.5\% of the
ISR $\psp \to \gamma \chico$ and $\gamma \chict$ background events, respectively.
 The residual yields of $\chico$ and $\chict$ events from ISR $\psp$ decays are expected to be $0.84\pm
0.15$ and $0.43\pm0.05$, respectively, where the uncertainties from intermediate branching
 fractions and $\psi(2S)$ production cross-section via ISR are considered. The selection efficiency of
 this requirement is 85.5\% for the $\chico$ signal and 80.9\% for the $\chict$ one.

A four-constraint (4C) kinematic fit constraining the four-momenta of the final-state particles to the initial $\EE$ collision system is performed. In $\EE\to \gamma \chicj$, an additional constraint is used, constraining the mass of the $\MM$ pair
to the $\jpsi$ nominal mass, giving a
five-constraint (5C) kinematic fit. Kinematic fits with $\chi^{2}_{\rm 5C}<25$ for $\EE\to\gamma
\chicj$ and $\chi^{2}_{\rm 4C}<30$ for $\EE\to \gamma \etac$ are required to improve the resolutions of
 the momenta of charged tracks and the energies of photons, and to suppress backgrounds with more
than two photons, such as ISR processes.

\begin{figure}[htbp]
\includegraphics[height=7cm,angle=-90]{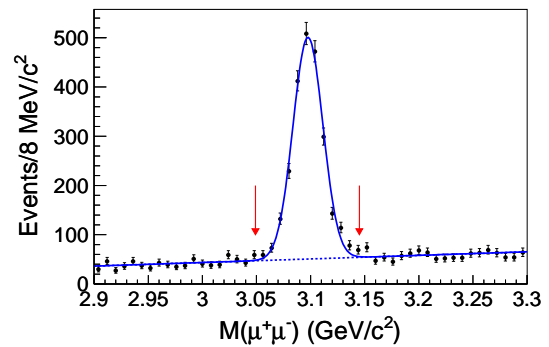}
\caption{The invariant mass distribution of $\MM$ pairs from
all the data samples before the application of the 5C kinematic fit. The solid curve is the fit, and the dotted line is the fitted
background. The arrows show the boundaries of the defined $\jpsi$ signal region.}
\label{mjpsi}
\end{figure}

The invariant mass distribution of the $\MM$ pair from the continuum, $\yfos$, and $\yfis$ data
samples, prior to the application of the 5C fit, is shown in Fig.~\ref{mjpsi}, together with the result of fitting the data to the sum of a Gaussian function for the $J/\psi$ and a first order polynomial for the background.  In the plot, a clear $J/\psi$ signal is observed.
We define the $\jpsi$ signal region as $|M_{\MM}-m_{\jpsi}|<48~\mevcs$ corresponding to three times the detector resolution, where $m_{\jpsi}$ is the $\jpsi$
mass~\cite{C38.090001}.

After all of the above requirements, some non-peaking background events
are observed in the processes $e^+e^- \to \gamma \chicj$ and $\gamma\etac$ at the studied C.M. energy points.

\begin{figure}[hbp]
\begin{center}
\psfig{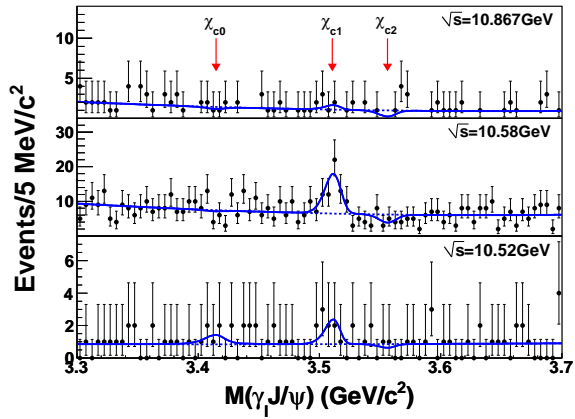}
\end{center}
\caption{The $\gl\jpsi$ invariant mass spectra at $\sqrt{s}$ = 10.52 (bottom), 10.58 (middle), and 10.867~GeV (top) together with fit results. The
points with error bars show the data and the solid curves are the fit functions; the dashed curves show the fitted backgrounds contributions. The arrows show the expected peak positions for the $\chicz$, $\chico$, and $\chict$ states.}
\label{data}
\end{figure}

Figure~\ref{data} shows the $\gl\jpsi$ invariant mass distributions for the data. A clear $\chico$ signal is observed in the $\yfos$ data sample, but is not evident in the other two data samples. Unbinned extended maximum-likelihood fits to the $M_{\gl\jpsi}$ distributions are performed to extract the $\chicj$ signal yields.
The $\chicj$ signal shapes in the fits are a Breit-Wigner (BW) function convolved with a Log-normal~\cite{novo} function with all the values of the $\chi_{cJ}$ resonance parameters fixed from the fits to MC simulations; second-order polynomial functions are used to describe the background distributions. The MC-simulated $\chicj$ signals have mass resolutions around
$6~\mevcs$ with small low-mass tails due to the measurement of $E(\gl)$. The results from the
 fits are listed in Table~\ref{sum}. The statistical significances of the $\chicj$ signals are calculated
using $\sqrt{-2\ln(\mathcal{L}_0/\mathcal{L}_{\rm max})}$, where $\mathcal{L}_0$ and
$\mathcal{L}_{\rm max}$ are the maximized likelihoods of the fits without and with the $\chicj$ signal,
respectively. The statistical significance of the $\chico$ signal in the $\yfos$ sample is $5.2\sigma$. The signal significance remains at $5.1\sigma$ when convolving the likelihood profile with a Gaussian function of width equal to the total systematic uncertainty discussed below.  The $\chicj$ signals in the continuum sample and the $\yfis$ sample are not
significant, as indicated in Table~\ref{sum}.

\begin{figure}[hbp]
\begin{center}
\psfig{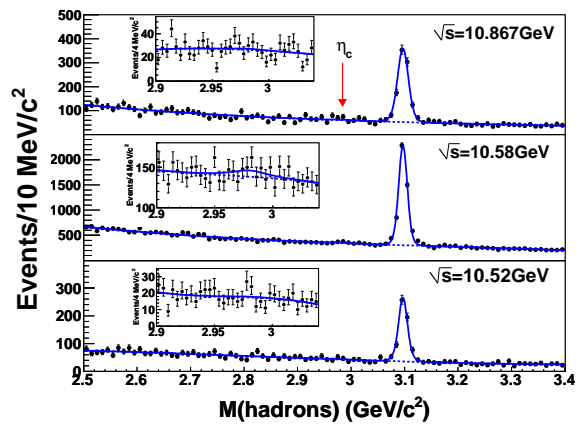}
\end{center}
\caption{The mass distributions for the sum of the five $\etac$ decay modes at $\sqrt{s}$ = 10.52 (bottom), 10.58 (middle), and 10.867~GeV (top).
The points with error bars show the data and the curves show the best-fit results; the dashed curves show the backgrounds contributions. The insets show the $\etac$ region. The $\jpsi$ signals are produced via ISR.}
\label{sumetac}
\end{figure}

Figure~\ref{sumetac} shows the $\etac$ invariant mass distributions for the five hadronic final states combined.
Clear signals resulting from the production of $J/\psi$ by ISR are present, while no significant $\etac$ signal is evident.


We perform a simultaneous fit to the five $\etac$ final states, in which the ratio of the yields in each channel is fixed to the ratio of
 $\BR_{i}\eff_{i}$, where $i$ is the $\etac$ decay-mode index, $\BR_{i}$ is the branching fraction taken from the Particle Data Group
(PDG)~\cite{C38.090001}, and $\eff_{i}$ is the reconstruction efficiency determined from MC
 simulation. In the fit, we use a BW function convolved with a Gaussian resolution
function to describe the $\etac$ signal; the values of all parameters are fixed from the fits to MC simulations. A Gaussian function with free parameters is used to describe the $\jpsi$ signal, and a second-order Chebyshev polynomial function is used for the backgrounds. The fit results are shown in Fig.~\ref{sumetac} and summarized in Table~\ref{sum}.

The Born cross section for $\EE \to \gamma X$ is given by
the formula
\beq\label{cross_section}
\sigma_{\rm B}(\EE \to \gamma X)=\frac{N^{\rm obs}\times|1-\prod|^{2}}{\mathcal{L} \times \sum_i\BR_i
\eff_i \times (1+\delta)_{\rm ISR}},
\eeq
where $N^{\rm obs}$ is the number of signal events obtained from the fit, $\mathcal{L}$ is the
integrated luminosity of the data sample, $\BR_i$ and $\eff_i$ are the branching fraction and the
 detection efficiency of the $i$-th $X$ decay mode ($\chicj$ is reconstructed in one decay mode and
$\etac$ in five decay modes). $(1+\delta)_{\rm ISR}$ is the radiative-correction factor, calculated using the formula given in Ref.~\cite{Fiz.41.733}, and
$|1-\prod|^{2}$ is the vacuum polarization factor, calculated according to Ref.~\cite{J.C.66.585}.
 The obtained Born cross sections for $\EE \to \gamma \chicj$ and $\gamma\etac$ are listed
in Table~\ref{sum} together with all the parameter results needed for the cross section calculation.


For all processes but $\EE\to\gamma\chico$ at $\sqrt{s}=10.58~\gev$, upper limits at 90\% credibility level (C.L.)~\cite{CL} on the numbers of signal events ($N^{\rm UL}$) and the Born cross sections ($\sigma_B^{\rm UL}$) are determined by solving the equation
\beq\label{cross_section}
\frac{\int^{x^{\rm UL}}_0 \mathcal{F}_{\rm likelihood}(x)dx}{\int^{+\infty}_0\mathcal{F}_{\rm likelihood}(x)dx} = 90\%,
\eeq
where $x$ is the assumed signal yield or Born cross section, and $\mathcal{F}_{\rm likelihood}(x)$ is the corresponding maximized likelihood from a fit to the data.
To take into account the systematic uncertainties discussed below, the likelihood is convolved with a Gaussian function whose width equals the corresponding systematic uncertainty.

\linespread{1.5}
\begin{table*}[htbp]
\caption{
Measurements of $\EE \to \gamma \chicj$ and $\EE \to \gamma
\etac$ at $\sqrt{s} = 10.52~\gev$, $10.58~\gev$, and $10.867~\gev$. $\eff$(\%) represents efficiency for the $\EE \to \gamma \chi_{cJ}$, and value of $\Sigma_{i}\BR_{i}\eff_{i}$ for the $\EE \to \gamma \etac$. $\Sigma(\sigma)$ is the statistical signal significance; $\sigma_{\rm syst}$(\%) is the systematic
uncertainty on $\sigma_{\rm B}$. The Born cross sections are given with statistical (first) and systematic (second) uncertainties.
}\label{sum}
\small
\begin{tabular}{c | c | c | c | c | c | c | c | c | c | c}
\hline Channel &$\sqrt{s}~(\gev)$ & $N^{\rm obs}$ & $N^{\rm UL}$ & $\Sigma(\sigma)$ & $\eff$(\%) & $|1-\prod|^{2}$ & $(1+\delta)_{\rm ISR}$ & $\sigma_{\rm syst}$(\%) & $\sigma_{\rm B}~(\rm fb)$ & $\sigma_{\rm B}^{\rm UL}~(\rm fb)$ \\
\hline
$\EE \to \gamma \chicz$ &\multirow{4}{*}{10.52}&$2.9^{+4.0}_{-3.3}$  & 9.6  & 0.9 & 19.0 &\multirow{4}{*}{0.931}  & 0.732868 &10.7& $286.2^{+394.7}_{-325.6}\pm{30.7}$ & 957.2 \\
$\EE \to \gamma \chico$ &&$4.8^{+3.6}_{-2.9}$ & 10.4 & 1.9 &20.8  &  &0.733432 &8.9 &$16.2^{+12.1}_{-9.8}\pm{1.4}$ &34.9\\
$\EE \to \gamma \chict$ &&$-0.8^{+2.3}_{-1.6}$  &4.5  & - &19.9  &  &0.733675 &12.8&$-5.0^{+14.3}_{-10.0}\pm{0.6}$ &28.9\\
$\EE \to \gamma \etac$ &&$6.8^{+14.8}_{-14.3}$  &30.8 & 0.5 &0.79  & &0.732788  &11.3 &$9.0^{+19.5}_{-18.8}\pm{1.0}$&40.6 \\\hline
$\EE \to \gamma \chicz$ &\multirow{4}{*}{10.58}&$-1.6^{+9.8}_{-8.9}$  &16.5 & - &18.9 &\multirow{4}{*}{0.930}  &0.732725 &13.1&$-20.0^{+122.3}_{-111.0}\pm{2.6}$ &205.9 \\
$\EE \to \gamma \chico$ &&$39.0^{+9.5}_{-8.8}$  &- &5.2 &19.9  &  &0.73329 &10.0&$17.3^{+4.2}_{-3.9}\pm{1.7}$ &-\\
$\EE \to \gamma \chict$ &&$-8.7^{+5.7}_{-5.0}$  &7.2 & - &19.8  &  &0.733532  &20.9&$-6.8^{+4.5}_{-3.9}\pm{1.4}$ &5.7 \\
$\EE \to \gamma \etac$ &&$67.2^{+42.0}_{-39.2}$  &125.9 &1.8 &0.78  &  &0.732645  &13.0 &$11.3^{+7.0}_{-6.6}\pm{1.5}$&21.1\\\hline
$\EE \to \gamma \chicz$ &\multirow{4}{*}{10.867}&$-1.3^{+4.0}_{-3.2}$  &7.0 &- &17.7 &\multirow{4}{*}{0.929}  &0.732054  &9.4&$-101.4^{+312.0}_{-249.6}\pm{9.5}$ &543.7 \\
$\EE \to \gamma \chico$ &&$1.9^{+3.4}_{-2.6}$  &7.9 &0.7 &16.8  &  &0.73262 &13.4&$5.8^{+10.5}_{-8.0}\pm{0.8}$ &24.3\\
$\EE \to \gamma \chict$ &&$-2.8^{+3.2}_{-2.4}$  &5.3 & - &16.3  &  &0.732863 &14.4&$-15.7^{+17.9}_{-13.4}\pm{2.3}$ &30.3\\
$\EE \to \gamma \etac$ &&$12.3^{+18.2}_{-17.4}$  &42.3 &0.9 &0.76  &  &0.731974 &9.1 &$12.3^{+17.3}_{-18.1}\pm{1.1}$&42.2\\\hline
\end{tabular}
\end{table*}

\begin{figure}[htbp]
\begin{center}
\psfig{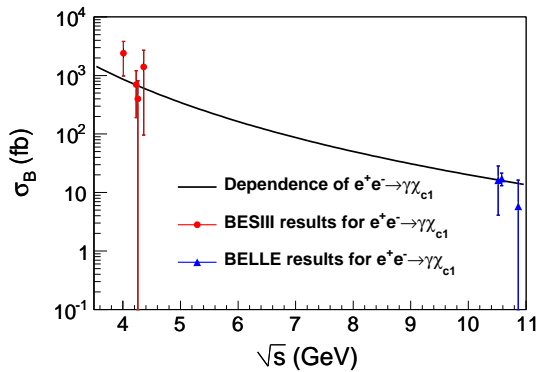}
\end{center}
\caption{Measured cross sections for $\EE\to\gamma\chico$ as a function of C.M. energy. Error bars contain both the
statistical and systematic uncertainties. The curve shows the result of fit with a function proportional to $1/s^{n}$.
}
\label{depend}
\end{figure}

Combining the measurement of $\sigma_{\rm B}(\EE\to\gamma\chico)$ from BESIII~\cite{C39.041001} and this analysis, we show the cross section as a function of $\sqrt{s}$ in Fig.~\ref{depend}.
We fit these data points with a function proportional to $1/s^{n}$ assuming that the reaction $e^+ e^- \to \gamma \chi_{c1}$ proceeds through the continuum process only: from a fit to the seven points for $\EE\to\gamma\chico$, we find $n = 2.1^{+0.3}_{-0.4}$.
The significance of the fitted $n$ is 2.2$\sigma$, calculated
using $\sqrt{\chi^2_0-\chi^2_{\rm min}}=2.2$, where $\chi^2_0$ is the $\chi^2$ with $n$ fixed at 0,
and $\chi^2_{\rm min}$ is the minimum $\chi^2$ with the value of $n$ free, respectively.
Adding an additional possible resonance, such as $\psi(4040)$, $\psi(4160)$, $Y(4260)$, or $\Upsilon(4S)$, the largest change in the fitted value of $n$
is 0.3. The result is consistent with the prediction by NRQCD with all leading relativistic corrections included in Ref.~\cite{1712.06165}. Due to the large uncertainties, we do not fit the $\sqrt{s}$ dependence of $\EE \to \gamma \chicz$, $\EE \to \gamma \chict$, or $\EE \to \gamma \etac$.

\renewcommand{\baselinestretch}{1.1}


There are several sources of systematic uncertainty in the Born cross section measurements,
including detection efficiency, the statistical error of the MC efficiency, trigger simulation, intermediate state
branching fractions, resonance parameters, the distribution of $\theta_\gamma$ for
$\EE\to \gamma \chict$, fit uncertainty, the $s$ dependence of the cross sections, and the integrated luminosity.
The systematic uncertainty for detection efficiency is a final-state-dependent combined uncertainty for all the different types of particles detected, including tracking efficiency, PID, $\ks$ selection, and photon reconstruction.

Based on a study of $D^{\ast+}\to D^0(\to K_S^0\pi^+\pi^-)\pi^+$, the uncertainty in tracking efficiency is taken to be 0.35\% per
track. The uncertainties in PID are studied via $\gamma\gamma\to\LL$ for leptons and a low-background sample of $D^*$ decay for charged kaons and pions. The studies show uncertainties of 2.2\%
for each muon, 1.0\% for each charged kaon, and 1.2\% for each charged pion.

Comparison of the $\ks$ selection efficiencies determined from data and MC results in $1 - \frac{\eff_{\rm data}}{\eff_{\rm MC}} = (1.4 \pm
0.3)$\%; 1.7\% is taken as a conservative systematic uncertainty. The
uncertainty in the photon reconstruction is 2.0\% per photon, according to a study of radiative Bhabha
 events. For each final state, the final detection efficiency uncertainty is obtained by adding all
sources in quadrature.

The statistical uncertainty in the determination of efficiency from MC is less than 1.0\%. We include uncertainties
 of 4.8\% and 0.6\% from trigger simulations for $\EE \to \gamma \chicj$ and $\gamma
\etac$, respectively. The uncertainties from the intermediate decay branching fractions are taken from
Ref.~\cite{C38.090001}. For $\EE \to \gamma \chicj$, the total uncertainties from the branching
fractions are obtained by adding all relative uncertainties in quadrature. For $\EE \to \gamma
\etac$, the total uncertainty from the branching fraction is obtained by summing in quadrature over the five decay
 modes with weight factors equal to the corresponding efficiency. The uncertainties from the resonance
 parameters are estimated by changing the values of mass and width of a resonance by 1$\sigma$ in the
fits~\cite{C38.090001}. Additionally, for the mode $\EE\to \gamma \chict$, the uncertainty from
simulating the $\theta_\gamma$ dependence is estimated to be 8.2\% by comparing the difference between a phase space distribution and the angular distributions of $(1 \pm {\rm cos}^2\theta_{\gamma})$.

In determining the number of signal events from the fits to data, the fit range and the choice of the
 function to describe the backgrounds are the main sources of systematic uncertainty. For the
latter, the background shapes are replaced by an exponential form or a higher-order Chebyshev polynomial, and the largest difference compared to the nominal fit result is taken as the related systematic uncertainty.
 Changing the $s$ dependence of the cross sections from fitted values of $n$ to a large number, e.g., $n=4$, gives very small differences in the radiative-correction factor ($<1\%$).
The total luminosity is determined to
 1.4\% precision using wide-angle Bhabha scattering events. All the uncertainties are summarized in
Table~\ref{systematic} and, assuming
all the sources are independent, summed in quadrature for the total systematic uncertainties.

\begin{table*}[htbp]
\begin{threeparttable}
\caption{Relative systematic uncertainties (\%) in the cross-section measurements for $\EE \to \gamma\chicj$ and
$\gamma \etac$ at $\sqrt{s} = 10.52~\gev$, $10.58~\gev$, and
$10.867~\gev$. When three values are given in a cell, they apply to $\chicz$, $\chico$, and $\chict$, respectively; otherwise a single number applies to all states.}
\label{systematic}
\begin{tabular}{c | c c c | c c c}
\hline
Final state & \multicolumn{3}{c|}{$\gamma\chi_{c0}/\chi_{c1}/\chi_{c2}$} &  \multicolumn{3}{c}{$\gamma\etac$} \\\hline
C.M. energy ($\gev$)  &  10.52 & 10.58 & 10.867 & 10.52 & 10.58 & 10.867 \\\hline
Detection efficiency     & 6.0 & 6.0 & 6.0 & 2.8 & 2.9 & 3.0 \\
MC sample size & 1.0 & 1.0 & 1.0 & 1.0 & 1.0 & 1.0 \\
Trigger & 4.8 & 4.8 & 4.8 & 0.6 & 0.6 & 0.6 \\
Branching fractions & 4.8/3.7/3.8 & 4.8/3.7/3.8 & 4.8/3.7/3.8 & 7.5 & 7.6 & 7.7 \\
Resonance parameters &1.7/0.3/0.7 &2.0/0.2/0.1 &0.9/0.4/2.1 & 0.6 & 1.7 & 2.0 \\
$\theta_\gamma$ distribution & -/-/8.2 & -/-/8.2 & -/-/8.2 & - & - & - \\
Fit uncertainty &5.3/1.9/4.5 &9.1/5.0/17.1 &1.4/10.3/7.7 & 7.7 & 9.8 & 2.6 \\
Integrated luminosity & 1.4 & 1.4 & 1.4 & 1.4 & 1.4 & 1.4 \\\hline

Sum in quadrature & \quad 10.7/8.9/12.8 \quad & \quad 13.1/10.0/20.9 \quad & \quad 9.4/13.4/14.4
\quad & \quad 11.3 \quad & \quad 13.0 \quad & \quad 9.1 \quad \\\hline

\end{tabular}
\end{threeparttable}
\end{table*}

In summary, we perform measurements of $\EE \to \gamma \chicj$ $(J = 0, 1, 2)$ and $\gamma\etac$
 at $\sqrt{s} = 10.52~\gev$, $10.58~\gev$, and $10.867~\gev$ using a $921.9~\infb$ data sample
taken by the Belle detector. A clear $\chico$ signal is observed at 10.58~$\gev$ with a
statistical significance of $5.2\sigma$, and the Born cross section is measured to be
$(17.3^{+4.2}_{-3.9}(\rm stat.)\pm 1.7(\rm syst.))~\fb$. For the cases where a $\chicj$ or $\etac$ signal is not
evident, upper limits on the Born cross sections are determined at 90\% C.L. Using the cross
sections measured at three different $\sqrt{s}$ in this analysis and from BESIII at
much lower $\sqrt{s}$ and assuming the reaction $e^+ e^- \to \gamma \chi_{c1}$ proceeds through the continuum process only, we determine the cross section $s$-dependence to be $1/s^{2.1^{+0.3}_{-0.4}\pm0.3}$ for $\EE \to \gamma \chico$.

We thank the KEKB group for the excellent operation of the accelerator; the KEK cryogenics group
for the efficient operation of the solenoid; and the KEK computer group, the National Institute of
Informatics, and the Pacific Northwest National Laboratory (PNNL) Environmental Molecular Sciences
Laboratory (EMSL) computing group for valuable computing and Science Information NETwork 5 (SINET5)
network support.  We acknowledge support from the Ministry of Education, Culture, Sports, Science,
and Technology (MEXT) of Japan, the Japan Society for the Promotion of Science (JSPS), and the
Tau-Lepton Physics Research Center of Nagoya University; the Australian Research Council;
Austrian Science Fund under Grant No.~P 26794-N20; the National Natural Science Foundation of China
under Contracts No.~11435013, No.~11475187, No.~11521505, No.~11575017, No.~11675166, No.~11705209;
Key Research Program of Frontier Sciences, Chinese Academy of Sciences (CAS), Grant
No.~QYZDJ-SSW-SLH011; the  CAS Center for Excellence in Particle Physics (CCEPP);
the Ministry of Education, Youth and Sports of the Czech Republic under Contract No.~LTT17020;
the Carl Zeiss Foundation, the Deutsche Forschungsgemeinschaft, the Excellence Cluster Universe,
and the VolkswagenStiftung; the Department of Science and Technology of India; the Istituto
Nazionale di Fisica Nucleare of Italy; National Research Foundation (NRF) of Korea Grants
No.~2014R1A2A2A01005286, No.2015R1A2A2A01003280, No.~2015H1A2A1033649, No.~2016R1D1A1B01010135,
No.~2016K1A3A7A09005 603, No.~2016R1D1A1B02012900; Radiation Science Research Institute, Foreign
Large-size Research Facility Application Supporting project and the Global Science Experimental Data
Hub Center of the Korea Institute of Science and Technology Information; the Polish Ministry of
Science and Higher Education and the National Science Center;
the Ministry of Science and Higher Education and
Russian Science Foundation (MSHE and RSF), Grant
No. 18-12-00226 (Russia); the Slovenian Research
Agency; Ikerbasque, Basque Foundation for Science, Basque Government (No.~IT956-16) and
Ministry of Economy and Competitiveness (MINECO) (Juan de la Cierva), Spain; the Swiss National
Science Foundation; the Ministry of Education and the Ministry of Science and Technology of Taiwan;
and the United States Department of Energy and the National Science Foundation.

\renewcommand{\baselinestretch}{1.2}


\begin{thebibliography}{**}
\bibitem{conjugate} Charge-conjugate decays are implicitly assumed throughout the paper.
\bibitem{novo} The Log-normal function is defined as $f(x)=N_{0}\times{\rm exp}[-\frac{1}{2}({\rm ln}^{2}(1+\Lambda(x-x_{0}))/\tau^{2}+\tau^{2})]$ with $\Lambda={\rm sinh}(\tau\sqrt{{\rm ln} 4})/(\sigma\sqrt{{\rm  ln}4})$. The parameters represent the mean ($x_{0}$), the width ($\sigma$), and the tail asymmetry ($\tau$).
\bibitem{CL} In common high-energy physics usage, this Bayesian interval has been reported as the ``confidence interval,'' which is a frequentist-statistics term.
\end{thebibliography}
\end{document}